# Tribute to an astronomer: the work of Max Ernst on Wilhelm Tempel

Yaël Nazé (Research associate FNRS, ULg)

Abstract: In 1964-1974, the German artist Max Ernst created, with the help of two friends, a series of works (books, movie, paintings) related to the astronomer Wilhelm Tempel. Mixing actual texts by Tempel and artistic features, this series pays homage to the astronomer by recalling his life and discoveries. Moreover, the core of the project, the book *Maximiliana or the Illegal Practice of Astronomy*, actually depicts the way science works, making this artwork a most original tribute to a scientist.

## 1. Art and the astronomer

Amongst the many artworks related to astronomy, a subset is linked to astronomers themselves rather than celestial objects or theories (Soderlund 2010, Nazé 2015). These paintings or sculptures fall in two broad categories. The first comprises portraits intended to represent specific people. The likeness to the actual scientist depicted varies greatly from artwork to artwork. Some portraits bear a close resemblance to the person represented – usually, these are contemporaneous and the artist actually met the scientist (e.g. *Galileo* by Justus Sustermans in 1636). Others have only a remote resemblance to the model, which usually happens when portraits are made well after the subject's death. For example, the famous *School of Athens* by Raphaël, despite being a masterpiece, cannot be considered to show the actual faces of Plato, Aristotle, or Archimedes. In this context, plain errors are sometimes made: the portrait by Ford Madox Brown of *William Crabtree Observing the Transit of Venus*, for example, displays a man of great age while Crabtree was only 29 at the time of the event!

https://fr.wikipedia.org/wiki/L'Astronome_%28Vermeer%29#/media/File:VERMEER_-_El_astr%C3%B3nomo_%28Museo_del_Louvre,_1688%29.jpg

*Figure 1: The astronomer (c. 1668), by J. Vermeer, oil painting, 51 × 45 cm, Musée du Louvre, Paris.*

The second category of astronomers' portraits is composed of works depicting a generic figure unrelated to any specific scientist. In order for the astronomer to be recognizable, such portraits must display the typical attributes of that profession. These could be books, instruments or images of celestial objects, and the choice depends on epoch and geographical region under consideration. These "practical" attributes are often complemented by "spiritual" characteristics – the (always male) astronomer is a pillar of society, he is often lonely, not very young, and thoughtful (e.g. Fig. 1) – sometimes to the point of ridicule (e.g. in the series of satirical paintings *Phases de la Lune* and *Les Astronomes* by the Belgian painter Paul Delvaux).

There is, however, one artwork linked to an astronomer that does not belong to either of these two categories. Yet, it is probably the most imaginative and interesting tribute to an astronomer: *Maximiliana*, by Max Ernst, and its associated art pieces. This series goes well beyond showing the physiognomy of an astronomer; instead, it trys to convey the actual work and life of one of them. The choice of subject, Wilhelm Tempel, was not made at random. This article begins with a section summarizing the life of both Tempel and Ernst, as well as their common points. It continues with a presentation, for the first time, of the full extent of the *Maximiliana* project, with details on each artwork and its interpretation.

*2. The two main protagonists*

*2.1 Wilhelm Tempel (1821-1889)*

Born into a poor family from Saxony, Ernst Wilhelm Leberecht Tempel received only the most rudimentary education. He became a lithographer, and travelled throughout Europe between 1841 and 1857, notably spending three years in Copenhagen. Because of a genuine interest in astronomy, he offered his skills at several observatories. However, at the time, his lack of academic qualifications and (recognized) experience in the field rendered his efforts largely futile (only a few short stays in Marseille and Bologna were possible). In 1858-1859, he spent time in Italy where he finally bought a refracting telescope with his own money and began observing from an open staircase attached to a Venetian palace, the *Scala del Bovolo*. After the discovery of one comet and one nebula in 1859, he moved to Marseille, where he was employed at the observatory from 1860-1861. He resumed his work as a lithographer during the following decade, observing from his home at night. When forced to leave France during the Franco-Prussian War of 1870, he moved back to Italy. He worked first at the observatory of Brera (1871-1875) and then in Arcetri (1875-1889), where he stayed until his death (for reviews on Tempel's life, see Dreyer 1889, Schiaparelli 1889, Iliazd 1964, Clausnitzer 1989, Bianchi et al. 2010 and references therein).

Tempel was a keen observer, and his acute eyesight led him to make the first observations of several objects, including 5 asteroids, 13 comets[1], and tens of nebulae. Several controversies are linked to Tempel: the first arose when he announced the discovery of a nebula near Merope in the Pleiades, since it could not be found by other astronomers using larger telescopes. In fact, this very faint nebulous feature could only be seen at low magnifying power, rendering its detection difficult – if not impossible – with the larger telescopes routinely used at high magnification in professional observatories. This led many well-known astronomers to doubt the nebula detection, with the result of Tempel having to defend himself publically (see e.g. Tempel 1880 and references therein).

A second controversy arose around the names of the first two asteroids he discovered. When discovering a new "planet", Tempel usually offered naming opportunities to others: the director of Marseille observatory, B. Valz, chose the name Angelina for asteroid (64) because of the astronomical work made in Notre Dame des Anges, near Marseille (Tempel 1861a); C.A. von Steinheil chose the name Maximiliana for asteroid (65) in honor of his patron, Maximilian II of Bavaria (Tempel 1861b); K. von

---

[1] 1859I, 1860IV, 1863IV, 1864II, 1869II, 1871II, 1871IV, 1871 VI, 1877V, and the four periodic 1866I (=55P Tempel-Tuttle, the source of the Leonid meteor shower), 1867II (=9P Tempel 1, visited by the Deep Impact and Stardust spacecrafts), 1869III (=11P Tempel-Swift-Linear), and 1873II (=10P Tempel 2). Note that Flammarion (1874) does not list comets 1871VI and 1877V, while Bianchi et al. (2010) do not list comet 1877V as it only deals with discoveries made with Tempel's own telescope. In addition, Tempel discovered "within 20 minutes" of Winnecke the comet 1870I, for which he could therefore be considered as co-discoverer. He also observed many other comets, for which he reported observations, notably in Astronomische Nachrichten (1860-1889).

Littrow chose the name Galatea for asteroid (74) (Tempel 1862a,b); C.A.F. Peters chose the name Terpsichore for asteroid (81) (Tempel 1865); and the Société Impériale des Sciences Naturelles de Cherbourg chose the name Clotho[2] for asteroid (97) (Tempel 1868). It was the choice of Angelina and Maximiliana in 1861 that triggered a debate.

While these names did not follow the common usage of choosing names from Greco-Roman mythology, they were certainly not the first ones to break the rule. In 1801, the first asteroid was originally named "Ceres Ferdinandea" by its discoverer G. Piazzi, to honour both Sicily, where the discovery was made (cereals are an important product of the island, hence its usual association with the goddess), and King Ferdinand III, whose patronage had made it possible. This choice led to some debate, with several alternative names proposed: Cupido, Cybele, Juno, Piazzi's planet, Ferdinandeam Sidus (following Herschel's name of Georgium Sidus for Uranus), Vulcan and Hera (Manara 1997). However, the controversy died down quickly, when Piazzi advocated for his right, as first observer, to name his discovery; moreover, usage soon led to the abandonment of the Ferdinandea epithet, essentially negating the original problem. In 1850, the 12[th] asteroid was named Victoria by its discoverer J.R. Hind, to honour both the Queen of England and the goddess of victory (Schmadel 2012 and references therein). This homage to a living ruler, again a departure from "tradition", was especially not welcomed in the United States, leading to a short but heated debate in the first volume of the Astronomical Journal (Gould 1850, Bond 1850, Hind 1850). The controversy again died down quickly, as the discoverer reminded that he had proposed two names, Victoria and Clio, and that the former also had some mythological links – hence it was kept. Finally, in subsequent years, asteroids (20) Massalia, (21) Lutetia, (25) Phocaea, (51) Nemausa, and (63) Ausonia were named after cities or regions, while the 45[th] asteroid was named Eugenia to honour the wife of Napoleon III and the 54[th] asteroid Alexandra to honour Alexander von Humbolt (Schmadel 2012). In all these cases, there was no 'fig-leaf' of a remote connection to Greco-Roman mythology but the unconventional designations were readily accepted without controversy.

The fierce attacks (e.g. Luther 1861, Hind 1861, Chambers 1866) on Tempel's asteroids only a few years later are thus surprising, especially considering that most came from Germany (the only support for Tempel's choices came from von Steinheil, 1861) and that German astronomers had avoided direct or strong objections to the previous unconventional choices. In any case, Cybele was proposed in Germany as an alternative to Maximiliana (Berliner Jahrbuch 1861) and this was rapidly adopted in the country: Astronomische Nachrichten lists Maximiliana in volume 55, Maximiliana and Cybele in volumes

---

[2] Tempel discovered the asteroid on February 17 and wrote on March 9 to Astronomische Nachrichten to announce the name chosen by the Cherbourg Society (Tempel 1868). However, the notes of the March 13 meeting of the Society mention the announcement of both the discovery by Tempel, corresponding member of the Society, and its naming offer to the Society. They continue with the choice of Clotho, following a proposal by the bureau members: as the decision was officially taken *after* Tempel's letter was sent to Astronomische Nachrichten, one can suppose that Tempel himself actually proposed the Parca name. In the same letter to Astronomische Nachrichten, Tempel proposed to use the names of the other two Parcae for asteroids (98) and (99), but this recommendation was not followed – these names were later used for asteroids (120) and (273)

58, 59, 61, 63, 64, but Cybele only in volumes 60, 62, 66, and 68 onward[3]. Cybele remains the official name of the 65[th] asteroid. The name of Angelina was kept for the 64[th] asteroid, though it was also challenged and Tempel himself proposed dropping the second 'n' to make it better agree with mythology (Tempel 1868). After 1870, only a few years later, the tradition of using Greco-Roman names for asteroids was completely abandoned and names were chosen at the whim of the discoverer, e.g. to honour their mentors, daughters or wives. In 1932, Asteroid (1217) was named Maximiliana to honour the work of the astronomer Max Wolf who had just died, while asteroid (3808) was named in 1982 in honour of Tempel himself, about a century after his death.

A last controversy arose around nebulae. After reports of nebular variability, Tempel aggressively defended their constant nature. First, considering all nebulae to be unresolved groups of stars, a variation would then only be explained if all stars varied simultaneously and in the same way, a possibility that Tempel dismissed as unphysical (Tempel 1863). Second, as he had observed himself, variable instrumentation, observer's abilities and assumptions, as well as weather conditions, affect the way nebulae are seen and, hence, are drawn. This leads to apparent differences in reports and to false conclusions if such factors are not taken into account (Tempel 1863, Tempel 1885, Bianchi et al. 2010). In making this argument, his patience as an observer (notably his regular habit of using several oculars to observe the same patch of the sky) and his experience as a lithographer and skilled artist played an important role, though these attributes were often underrated in professional astronomical circles. For similar reasons, Tempel also doubted reports of the detection of spiral structures in nebulae – although this time, he was in error. For astronomers that placed (somewhat blind) trust in their drawings and those of others, Tempel's ideas often seemed obstructive or irrelevant, and heated exchanges can be found in the literature (Tempel 1877a,b, Tempel 1878, see also Nasim 2010 for a summary of discussions on spiral structures and Nasim 2011 for a trial by J. Herschel towards less subjective hand drawings).

Finally, Tempel's reputation, status and treatment must also be noted. On the one hand, he received support from important astronomers such as G. Schiaparelli,;he won several prizes[4], and he had laudatory obituaries (Dreyer 1889, Schiaparelli 1889). On the other hand, there are several pieces of evidence showing that Tempel was not held in high regard, with some astronomers considering him at best as a "simple" amateur (Flammarion 1874; Iliazd 1964 – e.g. entry for year 1867 mentions Le Verrier declaring him "stupid"). As an illustration, although he was actually acting as the director of the Arcetri Observatory, he was never officially appointed and the job was left vacant rather than offered to him, something that caused him significant distress. In addition, his instruments were removed shortly before his death (Sawerthal 1889, Roberts 1889) and Tempel's rather sensitive character[5] interpreted this as a personal attack, although the actual primary cause was the commencement of long-needed repair works (Gautier 1889). Combined with the multiple scientific controversies, this perhaps explains why Tempel sometimes thought painters could better understand him than professional scientists ("*Je crois qu'un peintre paysagiste, qui aurait l'habitude d'observer et de réfléchir sur ces choses, se rangerait à mon opinion plus facilement que des météorologistes qui, entourés aujourd'hui d'une foule d'instruments qui notent et enregistrent automatiquement, ont désappris à se server de leurs yeux*", Tempel 1883).

## 2.2 Max Ernst (1891-1976)

Born near Cologne, Max Ernst was from a middle-class family (for a biography see e.g. Russell 1967, Bischoff 1994). He had a strict father, who was a teacher of deaf and mute students, as well as an amateur painter. After attending secondary school in Brühl, Ernst studied philosophy at the University of Bonn in 1910, but soon dropped out to concentrate solely on art. An admirer of Nietzche and Stirner, the young Ernst developed a rebellious streak, opposing his father and the Establishment. This mindset was exacerbated by the onset of World War I (as for many of his young contemporaries). It is therefore perhaps no surprise that he joined the Nihilist Dada movement, founding a Dada group in Cologne in 1918-1920 with J.T. Baargeld. Through this, he not only organized exhibitions but also exhibited his own works in Germany and France, shocking the public, art critics, as well as mainstream artists: Ernst thus became a prominent figure in Avant-Garde art. In 1922, he moved to Paris where he joined the surrealist circle around André Breton. Surrealism – emphasizing dreams and subjectivity – was predominantly a literary movement at the time, with painting only a small add-on[6]. Although Ernst was a very important Dada artist and a pioneer of surrealist painting during the 1920s, he never fully belonged to the surrealist circle. In fact, Ernst always avoided being defined by any label, preferring to stay at the periphery of any established grouping. This did not help his day-to-day finances, but ensured that he maintained total freedom over his art. Little by little, he distanced from Breton's group, verbally after 1925 and more officially in 1938 when they asked him to disown his friend Eluard. Following the outbreak of World War II, he was blacklisted by the Nazi Regime and was arrested several times by the French authorities. He therefore decided to leave France in 1941, reaching the United States through Spain with the help of Peggy Guggenheim. Under her protection, he did not experience as many problems as many other expatriate artists. However, his art was not really appreciated in his new home and he sold few pieces. He returned to France in 1953 and officially took on French citizenship five years later. After winning a prize at the Venice biennale, a wave of interest towards his work began, and he remained at the forefront of the artistic scene until his death in 1976.

Ernst's work is characterised by a departure from classical art, themes and techniques, in order to produce a new kind of art. However, he adopted no single style and his works are eclectic: several expressionist paintings while a young man, Dada and Surrealist works in between the two World Wars, pieces on the edge of abstraction afterwards, but also several bas-relief and large sculptures. He used a

---

bearing them philosophically. He often kept himself brooding over injuries, and sometimes without any real cause" (Schiaparelli 1889)

[6] However, surrealist painting is now better known than surrealist literature, with famous artists such as Dali, Magritte or Ernst. Ernst was the major figure at first, but Dali and his theatrical behaviour became the most famous representant of the movement after the 1940s.

wide range of innovative techniques (such as collage, frottage, decalcomania, and dripping), and he inspired many contemporary artists. Three recurrent themes can be identified in his artworks: birds, forests, and astronomy, in several forms[7]. One of his first paintings was an expressionist Sun (*Landscape and Sun*, 1909) and he declared, "*The significance of suns, moons, constellations, nebulae, galaxies, and space as a whole outside the earth zone have steadily taken root during the last century in human consciousness as well as in my work, and will most probably remain there.*" (Ernst & Schamoni, 1967 movie and 1974 book). From childhood, Ernst had been an avid reader and full of curiosity. His interests were broad: from science to magic, alchemy, and psychoanalysis. He considered religious themes outdated in the context of modern art, and thought science could provide interesting insights on nature that could inspire the new breed of artists (Stokes 1980). Allusions to astronomy are thus not surprising, but *Maximiliana* clearly represents an apex in his interest.

*2.3 The reasons of the choice*

One can speculate about why Ernst first became interested in Tempel. There is of course an obvious resonance in names: "Ernst" is the first given name of the astronomer and the family name of the artist, while "Maximilian(a)" is the given name of the artist and the original name of Tempel's renamed asteroid (65). This linguistic coincidence[8] may have appealed initially to the artist, but as more and more details about Tempel's life were uncovered by his friend, the writer and artist Iliazd, Ernst developed a sense of kindred spirit with the amateur astronomer, which led him to consider a large project linked to the astronomer.

Several parallels can be drawn between the lives of the two men. They both lacked formal qualifications, which marked them – for life – as being of lower status in the eyes of the elite (and possibly in their own). They both had to leave Germany to find work and achieve recognition in their chosen field, but both also suffered difficulties in France, their adoptive country, during wars with Germany. They both experienced financial difficulty – until 1953 in the case of Ernst and, for Tempel, throughout his life. Similarly, official recognition came late in life for both men, who both battled against the establishment. Indeed, an echo of Tempel's resentment against his scientific colleagues (see section 2.1) can be found in the 1974 Maximiliana book: Ernst comments that his work "*isn't appreciated by the specialists of fine arts, culture, behavior, logic and morals. But it enchants [his] accomplices: poets, pataphysicians, and some illiterates*".

---

[7] Archetypal solar disk or annulus in empty skies (*Landscape with Sun 1909, Sea and Sun 1925, Sun wheel 1926, Grey Forest 1927, Sea, Sun, Earthquake 1931, Forrest and Sun 1932, The entire city 1935, The phases of the night 1946, Humboldt current 1951/2, Forrest 1956, Sunspots 1960, The cardinals are dying 1962, The marriage of Earth and Sky 1962, Violet Sun 1962, Astral Hemisphere 1963, Yellow Sea, Blue Sun 1964, Reasonable Earthquake 1964, A moon is good thing 1970*), eclipse diagrams (*Of this men shall know nothing 1923*), constellations (the dada work *Fruit of a Long Experience 1919, Le grand Ignorant 1965*), planets, archetypal stars, nebulosities, and/or space-like trajectories (*Mon petit Mont Blanc 1922, The Bewildered Planet 1942, Birth of a Galaxy 1969, the Configurations* series in 1974, the lithographies *Terre des nébuleuses in 1975*). Ernst often referred to this astronomical imagery under the name "telluric images".

[8] It is graphically recalled in the 1964 book *Maximiliana* on page 4, which begins on the upper left by "Max" (beginning of Maximiliana) while "Ernst" (given name of the astronomer) appears at the upper right.

Finally and most importantly, they both put emphasis on the "art of seeing". In the 19<sup>th</sup> century, Tempel declared, "*Just as the human memory is less cultivated and exercised, owing to the mass of literature accumulated in the course of centuries, so the art of seeing truly is now being lost by the variety of instruments and artificial aids to vision*" (Tempel 1878, see also the end of Sect. 2.1). Tempel despised the fact that many professional astronomers lacked critical thinking, and seemed in thrall to the immediacy of answers provided by big instruments (one can speculate on his opinion of our obsession with computers and ready-to-use algorithms today!). For example, during the nebula controversies, he frequently reminded his "colleagues" that observing with several eyepieces and several telescopes, or in various weather conditions, could lead to different conclusions: the quick answer may not be the correct one, even if the instrument is ostensibly superior. This is a philosophy that, in itself, still remains true – and is nicely summarized by Ernst motto "*ce ne sont pas les grandes lunettes qui font les grands astronomes*" on page 25 of the 1964 book. Nonetheless, it should not be pushed to extremes, as demonstrated by Tempel's error with regard to spiral structures (see Nasim 2010). In Tempel's mind, secure conclusions could only be reached through patient and careful observing. In the 20<sup>th</sup> century, Ernst expresses similar sentiments: as a schoolboy, when asked to state his favourite activity, he readily answered "looking" (Ernst 1942, quoted in Russell 1967 and Stokes 1980). Moreover, Ernst, and more generally surrealists, wanted to make the unconscious visible, emphasizing the power of the (true) sight. Finally, it likely appealed to Ernst that Tempel, a scientist, placed the judgment of an artist over that of his colleagues (Tempel 1883, quoted in section 2.1).

It may be interesting to note that, in fact, there are also clear differences between Tempel and Ernst. First of all, their education level was very different, with Tempel only receiving the basics while Ernst received a classical education up to university level (even if he never graduated). Second, the financial troubles were of a different nature for the two men. On the one hand, Tempel came from a poor family, needed to work two jobs (lithographer during the day, astronomer at night) while in Marseille and ran a resourceless observatory in Arcetri, without funds for expenses or repair. On the other hand, Ernst came from a reasonably wealthy family, and most of his problems during his career came from his choice of remaining "on the edge". In addition, Ernst always had at least some level of fame in artistic circles from the 1910s onwards (with a first exhibition in 1912) and he was recognized as an inspiring figure amongst modern artists, while Tempel was clearly neither a leader nor an inspirational force of his time. Finally, both were rebels, but against very different things: Tempel fiercely fought innovations (large telescopes and use of photography), whereas Ernst battled against the old traditions… One could conclude that there were as many divergences as common points between the two men, but Ernst avoided the paradox by highlighting only the latter ones. He thus created a specific tribute to Tempel, in four parts, with the help from his friends Iliazd and P. Schamoni.

*3. The Maximiliana series*

*3.1 The prequel*

The *Maximiliana* project had a long gestation. Ernst's interest in Tempel dates back to before World War II, but it took some thirty years before he began to gather the necessary information for his 1964 book. Ernst probably came across asteroid (65) by chance, when looking for astronomical information or astronomical images (maybe in Flammarion's *Astronomie Populaire*, which was a widely used reference at the time for anyone interested in astronomy and where the 65[th] asteroid is quoted with the name Maximiliana, not Cybele). Ernst may then simply have been attracted by the name, which makes an amusing allusion typical of surrealist work. In any case it stimulated his initial interest in Tempel, which then never completely left his mind.

The first mention of Maximiliana in Ernst's work appeared in one of his "visible poems" of 1931 (Fig. 2, published again in 1948 in *A l'intérieur de la vue* with P. Eluard). This artwork displays the Sun as seen from the eight planets and from two asteroids, Maximiliana and Feronia. While the omission of Pluto, the – then – 9[th] planet, discovered the year before and largely publicized, might at first seem odd, this can be readily explained through the nature of the work: a collage. The main, background image comes from Camille Flammarion's *Astronomie populaire* (Chimirri 2009), and the French astronomer features all the planets known at the time of publication. Flammarion also chose to depict the Sun from two asteroids at the extremes of the main belt, although he avoided specifying these by name[9]. It is not known whether Ernst himself added the names for his artwork, or if they were already included in the copy he used, probably taken from one of the manuals and catalogs that he routinely used in his collages (see e.g. discussion of his method in Stokes 1980).

It should be noted that there is a big difference in tone between the 1964-1974 project and the earlier collage. The 1931 work, typical of surrealist mind, is clearly is making fun of scientific imagery by adding popular clichés to a serious scientific sketch: the large Sun seen from the nearby Mercury, with its grim-reaper figures, is a token of death; the solar disc associated with Venus features a suggestive breast typical of the goddess of beauty; that of Mars features a crab (typical of the Cancer constellation imagery) announcing the dangers of war; that of Jupiter, the most important God and planet, is reduced to an eye akin to the "divine eye" imagery; that of Earth has a table on fire symbolizing society in crisis; and those of the asteroids are faces, with a sinister death mask for Maximiliana – symbolizing the unruly artist. This collage is by no means a tribute to science; its ironic mood is typical of all Ernst's collages, including others with astronomical imagery (e.g. *Mon Petit Mont Blanc* where the planet Saturn is replaced by a female backside). However, irony is not a general characteristic of all his "celestial" artworks: for example, *The Bewildered Planet* or the major surrealist painting *Of This Men Shall Know Nothing* rather represent thoughtful reflexions on the nature of our world, a framework into which the 1964-1974 project fits. The 1931 collage thus is not so much interesting for its detailed imagery, but as direct evidence that Ernst knew Maximiliana, and Tempel, well before the 1960s.

---

[9] Names of (149) Méduse and (153) Hilda are mentioned as extremes in the 1890 text while in his book *Terres du Ciel* (1877), Camille Flammarion rather mentioned (8) Flore and (87) Sylvie as the closest and farthest asteroids, i.e. not (72) Feronia and (65) Maximiliana.

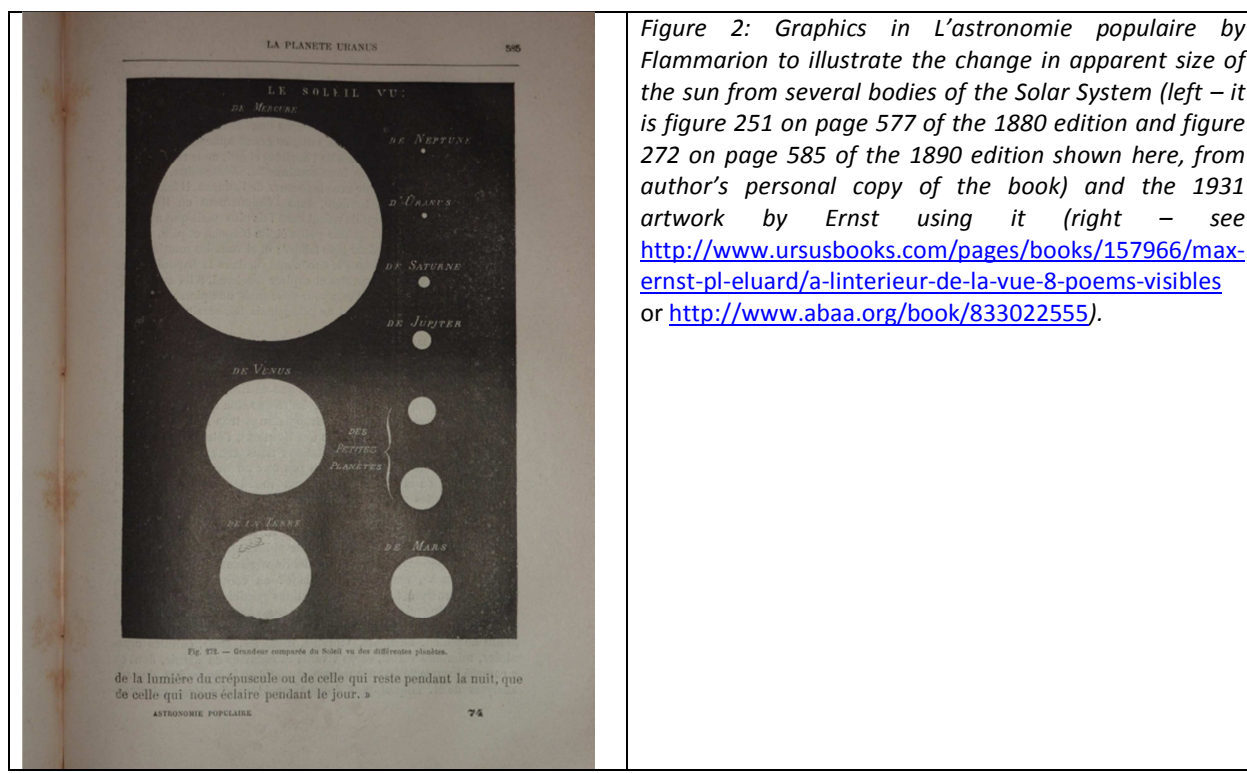

*Figure 2: Graphics in L'astronomie populaire by Flammarion to illustrate the change in apparent size of the sun from several bodies of the Solar System (left – it is figure 251 on page 577 of the 1880 edition and figure 272 on page 585 of the 1890 edition shown here, from author's personal copy of the book) and the 1931 artwork by Ernst using it (right – see* http://www.ursusbooks.com/pages/books/157966/max-ernst-pl-eluard/a-linterieur-de-la-vue-8-poems-visibles *or* http://www.abaa.org/book/833022555*).*

### 3.2 L'art de voir de Guillaume Tempel

*Maximiliana* revolves around knowledge of Tempel's life. However, the minutiae of Tempel's biography were poorly known in the early sixties. Uncovering these details became Iliazd's task. The poet specialised in "revivals", and had already brought to light several largely forgotten people such as Adrian de Monluc and Roch Grey (Greet 1982). He spent two years from 1962 to 1964 in France and Italy rooting through the archives of the observatories and cities where Tempel had lived and worked. He also read the scientific articles by and on Tempel. From these sources gathered, he wrote a 28-page, factual biography of Tempel, *L'art de voir de Guillaume Tempel* (Fig. 3) – one of the first detailed ones about Tempel. It is quite complete (though page numbers are missing in the bibliographic references) and contains only few errors[10]. However, despite the title, there is no discussion in Iliazd's book on the "art of seeing", only a list of facts and a short and quite romantic introduction summarizing in a few sentences the "martyrdom" of Tempel and his judgment on scientists (see quotes at the end of sections 2.1 and 2.3). Seventy copies of the book were published for an exhibition at the Point Cardinal, clearly as a preparation for the main book.

---

[10] Wrong volume number for the reference to The Observatory 1878 (it is volume 1 not 4), wrong roman number for some comets (e.g. 1863VI quoted as 1863V), wrong identification of 1867II as the first periodic comet of Tempel while it is his second one, and mistranslation or bad use of "telescope" from time to time (Tempel used refractors, not reflectors – the English language uses the word telescope for both instruments, but in French, one uses "lunette" for the former and "télescope" for the latter).

*3.3 Maximiliana ou l'exercice illegal de l'astronomie, the book*

The second step of the project is Ernst's masterpiece, the art book "*Maximiliana or the Illegal Practice of Astronomy*" (for previous discussions, see Greet 1982, Hubert 1984, and Spielman 1996). Composed of 30 pages[11], it was published in 1964 in a run of 75 copies (or to be precise 65+X copies, as mentioned on the last page). First, the reader encounters Tempel's words, deploring the loss of the art of seeing (see quote in section 2.3): this recalls the credo of both Ernst and Tempel, and gives some keys to understanding the book (see below). It is indeed not by chance that these words open the book.

Next, the title page lists the title and the three authors of the book (Tempel, Ernst, Iliazd) along with some dynamic hieroglyphic writing invented by Ernst for this work and two pairs of elegant, elongated figures. Each of these pairs is composed of the same drawing at different scales: a small black and larger red version form a kind of object-and-shadow pair that provide some perspective – 3D depth – to the scene. There is no obvious link between these drawings and astronomy in general or the asteroid (65) in particular, except that the figure on the rightmost side appears somewhat feminine and Maximiliana is a female name.

See http://art.famsf.org/search?search_api_views_fulltext=maximiliana for scans of the pages

*Figure 4: Representative excerpts of Maximiliana ou l'exercice illegal de l'astronomie (Ernst, Iliazd, Tempel, 1964), etchings, 50 × 37.5 cm. The top left panel shows page 3, a divider page with a bird-like drawing and some hieroglyphic writing invented by Ernst. Note that this text appears twice, at different scales, to provide a perspective effect. The top right panel shows page 4, an example of calligram with a sketch reminiscent of a night sky. The bottom panels display the two types of pages from the central part of the book: page 14 (left) mixes Tempel's poem with hieroglyphs and sketches by Ernst (both organized in four columns), while page 17 (right) reproduces published text by Tempel along with drawings by Ernst. Note the spiral-like features in both pages.*

Next comes the first the divider pages (see top left panel of Fig. 4 for an example). There are four of these pages in *Maximiliana* (pages 3, 9, 23, 29), to separate the three main "chapters" of the book. They share a similar structure: on the left side is an etching by Ernst with bird-like, elongated figures (without obvious link to astronomy), along with a sentence in Ernst's hieroglyphs, while the opposite side displays two copies of a vertically-arranged hieroglyphic text (again, one small in black, one larger with colours, to provide a 3D effect).

The rest of the book is split in three parts of unequal lengths (5, 13, and 5 pages), all containing astronomy-related text or images. In the first and last parts (see top right panel of Fig. 4 for an example), each page contains a calligram (i.e. a text arranged to create a visual image) and an etching. The calligrams are all different and are organized to make a grid, diagonals, horizontal or vertical waves,

---

vertical lines recalling rain, or small clusters of letters. These short texts were written by Ernst, except for the ones on page 7, 27, and 28 which are quotes of Tempel's words. They are:

- on page 4: *Maximiliana 65 planète située entre Mars et Jupiter fut découverte à onze heures le vendredi huit mars 1861 sur la terrasse de l'observatoire ancien de Marseille par Ernst Guillaume Leberecht Tempel* ;
- on page 5: *invisible à l'oeil nu, elle paraissait dans sa famille être la plus éloignée du Soleil* ;
- on page 6: *nommée ainsi, elle provoqua le grand mécontentement des mythologues fut passée sous silence et puis changée en Cybèle* ;
- on page 7: *depuis Pâques je porte dans ma poche l'annonce que mon télescope est à vendre*
  This text is an excerpt of a letter to Hummel, quoted in Iliazd (1964, entry for year 1865) and in Eichhorn (1961).
- on page 8: *soixante et onze ans plus tard ce nom tant décrié fut pris par la femme d'un astronome pour baptiser la planète 1217 découverte à l'aide de la photographie par son réputé mari.*
  It was indeed the wife of Max Wolf, co-discoverer of the asteroid with E. Delporte, who chose the name Maximiliana, but in fact it was Delporte who first proposed the object to be named in the honour of Wolf, who had just died (Schmadel 2012).
- on page 24: *tempéliennes 1859 I 1860 IV 1863 V* [should be IV] *1864 II 1866 I* [should be listed with the periodical cases] *1869 II 1871 II 1871 IV 1871 VI 1877 V les trois périodiques 1867 II 1869 III 1873 II il fut encore le premier à déceler le retour des comètes* ;
  The last sentence is erroneous, as Tempel was not the first astronomer to have calculated the return of a comet. It may have been used on purpose to enhance Tempel's aura, but it could also be a simple error by the artists coming from a lack of detailed astronomical knowledge and the (potentially confusing) fact that Tempel detected several periodic objects.
- on page 25: *la nébuleuse de Mérope une des Pléiades fut découverte le 19 octobre MCCCLIX* [should be MDCCCLIX] *sur l'escalier tournant du palais vénitien Contarini réalité d'abord contestée ce ne sont pas les grandes lunettes qui font les grands astronomes* ;
  The last part recalls one of Ernst's motto on his artworks: "*ce n'est pas la colle qui fait le collage*" (see Hubert 1984)
- on page 26: *Planètes massiliennes (64) Angelina 4 3 1861 observatoire (65) Maximiliana 8 3 1861 observatoire (74) Galatée 29 8 1862 10 Pythagore (81) Terpsichore 30 9 1864 26 Pythagore (97) Clotho 17 2 1868 26 Pythagore* ;
  This page lists the five asteroids discovered by Tempel, along with their discovery date, as reported by the astronomer, and the place of discovery (Marseille Observatory, or the address of Tempel's private houses, first at 10 Pythagore Street, then at 26 in the same street).
- on page 27: *il serait poétique de donner aux dernières planètes 97 98 et 99 les noms des 3 parques Clotho Lachesis et Atropos non pour couper le fil de la recherche pour clore la première centaine des petites planètes*
  This text is an excerpt from Tempel (1868)
- on page 28 : *le mie stelle mi hanno finora aiutato espero che mi aiuteranno ancora*
  This supposedly comes from a letter from Tempel to Silvani (mentioned with the date of January 24 1881 in Iliazd 1964) stored in Tempel Arcetri archives. However, the letter could not be found again in recent years (S. Bianchi, private communication).

The etchings in these two parts seem to mirror each other: the first drawing (p4 and 24) is dense and indistinct, recalling a sky full of stars and mysteries to be discovered; the second (p5 and 25) could be a celestial close-up (celestial spheres on p5, a more complicated pattern on p25); the third (p6 and 26) possesses six panels with figure-like sketches, probably suggesting astronomers at work; the fourth (p7 and 27) displays a double-star structure, an obvious reference to the sky but a distant allusion to Tempel's work (who did not observe this kind of object in detail); the last one (p8 and 28) alludes to discoveries of patterns or to the fabric of (curved) space-time. However, this mirror effect in images cannot be found in texts, which have clearly different tones: they sadly recall Maximiliana's naming drama in the first part while the main discoveries made by Tempel are cheerfully announced in the last part, which ends in a hopeful tone.

The central part alternates between two types of pages (see bottom of Fig. 4 for examples of each case). At the bottom of the first kind, on odd-numbered pages, are excerpts of actual articles by Tempel on clouds (pages 11, 13 and 15, from Tempel 1883), nebulae (page 17, from Tempel 1863), or aurorae (page 19, from Tempel 1871). The original scientific texts are often cut, avoiding details. They have been carefully chosen from Tempel's bibliography – indeed, all the texts are easy to understand by non-specialists and contain a certain artistic appeal, as they mostly concern discussion of colors and textures in sky phenomena. Moreover, they provide direct evidence of the originality of Tempel's ideas and discoveries, while underlining the opposition to Tempel's ideas from other scientists. In these pages, etchings can be found above the texts. By their position, they could be considered as direct representations of the texts. However, they are largely abstract in nature, with only birds and spirals being recognizable. They are thus not as direct illustrations as the images associated with the calligrams.

The second type of page in the central section, appearing on even pages, shows Tempel's melancholic poem "*der Glöckner*" at the top. Dated 1849, it recalls his childhood, when he was a bell ringer. While there is no direct link with astronomy (Tempel's astronomical activity really began a few years later), it may have been used by Ernst to create an atmosphere, illustrating the sadness of Tempel when faced with fierce attacks. The poem is associated with a complex hieroglyphic text (with some enlarged characters and/or small etchings): again, apart from a few archetypal stars and spiral-like features, there seems to be no direct link with astronomy.

The book finishes with the colophon, illustrated again by Ernst's hieroglyphs. In many ways, *Maximiliana* is a triptych: it is the result of the efforts of three men (Tempel, Iliazd, and Ernst); written in three languages (French, German, Italian); separated in three main parts; alluding to three types of objects (the controversial clouds, nebulae, and aurorae in the central part, the achievements in comet, asteroid, and nebula observations in the last part); and containing three types of text (scientific (excerpts of actual articles by Tempel); personal (excerpts of letters by Tempel); and literary/fictional) as well as three types of signs (common alphabet, Ernst's hieroglyphs, and Ernst's drawings).

At first sight, *Maximiliana* could be taken as the imaginary diary of an astronomer, a nice and relatively straightforward (hence not-so-impressive) homage to Tempel. On the other hand, it could also be considered as following a very classical adventure scenario: first the hero faces challenges (the first calligrams), then he demonstrates his value through painful episodes (the central part) and finally he overcomes all obstacles (the last calligrams). However, it goes well beyond that. In fact, its real aim is clearly stated from the start: it is a tribute to the "art of seeing" – *Maximiliana* ambitiously proposes to make one learn to *see*. To do so, the artists Iliazd and Ernst offer content in the form of progressive steps. First, there are Tempel's notes and poem, which are readily understandable because they are written in plain language and with the usual placement of letters. Second are the calligrams: these are written in plain language, but the position of the letters is unusual so that one needs some time to be able to make up words, and then sentences – one thus needs to begin practicing the "art of seeing", i.e. uncovering hidden knowledge. Finally, there are Ernst's hieroglyphs, which are novel and unprecedented – their actual meaning thus remains obscure, as we do not have the key to this graphic language. However, the argument is also double-edged. The actual meaning and importance of the "easy-to-read" text remains essentially hidden to the layman as it can only be understood in relation to Tempel's life and with some knowledge of astronomy and geophysics, as well as of history of science. On the other hand, the secret writing of Ernst displays recognizable shapes, hence gives the impression of being readable or at least of always being on the edge of readability.

The overall thesis of the artists, that seeing is an art, is thus clearly demonstrated: if one looks carefully enough, one can make sense of things that were apparently obscure at first, while things apparently easily understandable become worth a second look to gain a true, in-depth meaning. In addition, the mix between the different types of texts blurs the linear progression from readable to non-understandable that one would naively expect – as in actual scientific life, where some evidence may sometimes be easily accounted for, while other may be much more difficult to understand, without any specific temporal arrangement between the two. This variety of texts necessitates a non-linear, back-and-forth reading which also demonstrates that a global look may help put pieces together, enlightening the reader in the same way as the scientist in his/her work. Tempel, with his many discoveries, is provided as a successful example in this endeavour, which actually is typical of the way science works – the book thus also has a general character transcending specificities. Rarely do artworks demonstrate so efficiently the true nature of science, and it is in this respect that *Maximiliana* can be considered as one of the best tributes to a scientist, or to science in general. It may also be noted that the art of seeing implicitly assumes a human will be there to perform it – so this work is also a tribute to human activity, against the dehumanization engendered by technology.

### 3.4 Maximiliana oder die widerrechtlich Ausübung der Astronomie, the movie

Three years after the publication of the books, a short movie[12] (11m30s) was made with the same title as the "masterpiece" book, this time in the German language. The director was Peter Schamoni, who belonged to the New German cinema movement. During his career, Schamoni made several documentaries about artists, including five[13] with (or at least partially about) Max Ernst which demonstrates his close relationship with the artist. It must be noted that Schamoni considered Ernst as an archetype of the 20[th] century artist.

---

[12] It is now freely available on youtube https://www.youtube.com/watch?v=SEyqDp5b2ok
[13] Besides it, there are *Max Ernst – Entdeckungsfahrten ins Unbewusste* in 1963, *Was ist ein Wald? Günther Lüders spricht Max-Ernst-Texte* in 1974, *Dorothea Tanning: Insomnia* in 1978, and *Max Ernst – Mein Vagabundieren, meine Unruhe* in 1991.

This time, contrary to the two previous works, the piece does not begin with Tempel's opinion on the art of seeing but by Ernst shouting "*er hat genie, aber kein diplom*" from an open window. Portraits of Tempel – an original photograph as well as a surrealist portrait by Ernst – then appear. As the places where Tempel lived are successively displayed, Ernst reads the "*der Glöckner*" poem and tells Tempel's story, emphasizing the problems he had to face and the discoveries he made. Ernst then asserts that he has undergone the same difficulties in his life, even if this is not entirely true (see section 2.3). Short excerpts of scientific texts by Tempel as well as some more personal notes (those appearing in the 1964 book) are read. They are illustrated with Tempel's drawings as well as Ernst's imagery from the 1964 book[14]. Ernst concludes by stating that he feels exactly the same as the astronomer but in the domain of art: that the accumulation of knowledge and techniques has killed the faculty of truly seeing in fellow artists.

A peculiarity in the images is worth noting. When the commentary mentions the professional astronomers sceptical of Tempel, photographs of apparently eminent 19[th] century scientists are shown. However, contrary to expectations, these are not the portraits of Tempel's "enemies"; they are in fact not even related to astronomy (the French physician C.P. Robin, the French chemist E. Fremy and the French mathematician A.J.H. Vincent are notably shown)! In fact, Ernst here employs exactly the same technique as for his collages, using images that he had at hand and that fit well his subject. This is of course done on purpose, as photographs of many of Tempel's opponents actually exist: his choice underlines that the then famous academics "harassing" Tempel have now fallen into oblivion; their faces are not recognized today, contrary to his hero who has stood the test of time.

The movie is a nice complement to the 1964 book, but it does not achieve the same artistic impact. However, it unveils the second objective of Ernst with the *Maximiliana* project. Indeed, two protagonists can actually be distinguished in this movie: Tempel's drawings and Ernst's artworks as well as Tempel's notes and Ernst's texts are fully mixed, appearing successively or in parallel. Therefore, the whole movie is as a tribute to both men, the artist and the astronomer, rather than to the latter exclusively. It blurs even more the lines between them, to illustrate their common views, and one then begins to realize that *Maximiliana* is also a kind of autobiography for Ernst.

*3.5 Maximiliana, the second book*

In 1974, Schamoni's movie was extended through a book, again called *Maximiliana Or the Illegal Practice of Astronomy*. Different in both structure and aim to the 1964 book, this volume is much more of a traditional book on art (a square volume of 90 pages with a wide distribution) and not a true artwork. This time written in English, French, and German, it alternates texts and artworks without the intricate structure of the 1964 book. This book again throws another light on the project, reinforcing the clues from the movie, as it appears first and foremost concerned with Ernst. Indeed, it summarizes the artist's life and thoughts, presenting a large selection of his artworks (including poems and several astronomical paintings), and it ends with biographical notes on Ernst in a tabular form.

---

[14] A few errors can also be noted in this movie: dates of the scientific texts are incorrect (they should be December 1882 and not July 1878 for the Ciel & Terre reference and March 1878 rather than August 1881 for The Observatory reference); Maximiliana, asteroid (65), was not "taken over" by another astronomer – the name was simply (re)used for another asteroid, without any intentional attack on Tempel; Tempel's drawing of M51 is shown when the commentary mentions Merope's nebula; and Tempel is again presented as the first astronomer having calculated the orbit and periodicity of a comet.

This book contains much less detail about Tempel, with only few pages (7 out of 90) directly alluding to him or his work[15]. In particular, pages 78 and 79 display a photography of Tempel sitting side-by-side with Ernst wearing the same costume – a photomontage graphically illustrating the comparability of the two men. In parallel, a quote by the art historian and critic Werner Spies on page 75 sheds some light on the use of hieroglyphs for the 1964 book: "*It is possible to interpret this confrontation of script and stars. Just as now and then a star emerges from the host – as a moving planet, as a comet – so also does a cipher, whose ideogram becomes intelligible to us, rise up now and then from the heap of incomprehensibility. One could draw the conclusion that the limits of vision corresponds to the limits of understanding.*" This is in line with the interpretation given in section 3.3.

*3.6 The associated paintings*

Finally, Ernst also made a few paintings related to his *Maximiliana* project (Fig. 5): *Earth Seen from Maximiliana* (1963) and *The World of the Naïve* (1965).

The former painting is a direct echo to the "visible poem" of 1931 where Ernst first mentioned Maximiliana (sect. 3.1). However, there are several content and stylistic changes. This time the Sun is replaced by the Earth and the "planetary" discs resemble actual planetary surfaces, though their relative size has no direct scientific meaning anymore. Only four of the seven celestial bodies are identified: the first one is Maximiliana, of course; the second one is Venus and the next two are related to the Moon. These last two provide the only ironic touch to this generally "serious" painting: their names ("Lune couchée", "Lune debout") can be translated as the lying Moon and the standing Moon: obviously not scientific concepts. While its general layout is reminiscent of the 1931 collage, the style of this painting is much more akin to Ernst's contemporary works, such as the abstract *Configuration* series. All of them show dream-universes, where textures and colors abound.

The latter painting is much more directly related to the 1964 book, as it uses the same hieroglyphic writing. It is, however, atypical of Ernst's paintings as he made only one similar piece, *The World of Confusions*[16], also in 1965. Its structure is quite intricate. At the top, there appears to be a photograph of a stellar cluster, while two discs akin to planets with realistic surfaces can be found below. The rest of the painting contains panels full of Ernst's hieroglyphs, arranged to give the impression of 3D space. This organization resembles that found in typical sketches from popular science books where spacetime is presented. This painting thus represents a vivid portrait of our complex universe. However, the presence of the hieroglyphs also graphically states that the universe awaits our "reading" – for those who have not lost the "art of seeing". It thus directly resonates with the 1964 *Maximiliana* book, and in fact constitutes a graphical summary of it. In this context, the title may appear strange but it should be remembered that Ernst considered Tempel and himself as being amongst the naïve – as opposed to members of the elite or the "professionals" (see Section 2.3). In his view, naïve does not have the negative connotations as per use in common language: rather, it refers to those with the potential to unveil cosmos' mysteries but who are often under-rated.

---

[15] Pages 1 shows an extract of Tempel (1882), stating an incorrect date (May 1879); this article contains factual descriptions of many nebulae, but the description shown here, the one associated to object #2102, is particularly vivid and aesthetically-minded (p230-231), explaining its choice. Page 82 shows a photograph of Tempel and of his tomb, as well as an excerpt of the list of discoveries he published in 1868 (and which was later copied in Flammarion 1874). The next one presents again an excerpt of Tempel (1878) on the loss of art of seeing (see section 2.3), again with an incorrect date (19/7/1878 rather than 13/3/1878) while page 84 provides a summary of Tempel's life, actually a short version of the movie commentary. Two drawings of nebulae by Tempel are also shown on the page facing page 1 and on page 83 (those are Orion nebula and Lagoon nebula, respectively, and they correspond to Tavole 16 and 13 of Tempel's catalog kept in Arcetri archives).

[16] The French title is much more explicit regarding Ernst's intent: *Le monde des flous* (which could be translated as "world of blurred things") is a word game with *monde de fous* (crazy world).

---

*Figure 5: Left: Earth seen from Maximiliana (1963), oil on plexiglas, 35 x 27 cm, Theo Wormland Foundation, Bayerische Staatsgemäldesammlung, Munich; http://www.vontobel-art.com/fr/Ernst-Max/La-terre-vue-de-la-Maximiliana-aid55001 or http://willemsconsultants.hautetfort.com/archive/2009/12/21/l-art-de-voir.html*

*Right: The world of the naïve (1965); oil painting, 116,5 × 89,5 cm, Centre Pompidou, Musée national d'art moderne, Paris. https://www.centrepompidou.fr/cpv/resource/cEnzebL/rGE9eBB*

---

*4. Conclusion*

Over a decade from 1964-1974, Max Ernst, with the help of the poet Iliazd and the movie director Peter Schamoni, produced several works linked to the astronomer Wilhelm Tempel. The core of this effort was the production of one of the first biographies of Tempel (by artists rather than historians or astronomers) and a masterpiece artbook entitled *Maximiliana ou l'exercice illégal de l'astronomie*. The former was actually made in preparation for the latter, as a two-step function: first, it unveiled details on Tempel's life that gave Ernst a feeling of kindred spiritship with the astronomer, without which no further artwork would have been produced; second, its production led to the selection of events and texts that were then used in the rest of the series. These features summarize Tempel's discoveries and "martyrdom" (which was somewhat exaggerated), but also testify to his keen sight and his genuine interest in art. It should be remembered that these artists, through their work, have certainly participated to keeping alive the remembrance of Tempel, a poorly known astronomer up to that point.

The aim of the 1964 book is far-reaching, going well beyond a "simple" graphical biography or a fantasy astronomer's notebook. In fact, its reader learns how to make unintelligible things gradually comprehensible and how to take a broader look at things apparently understood, two things that scientists do in their daily work to increase knowledge. This artwork therefore certainly offers a much deeper tribute to science than classical portraits of astronomers. This is not the only objective of the work, as revealed by the other complementary pieces of the project (short movie, 1974 book and paintings): *Maximiliana* is also an autobiography of Ernst under the mask of Tempel's name. They had indeed experienced similar situations: absence of qualifications, exile, poverty and slights from the establishment. While several divergences between the two men also exist, these common points led Ernst to make Tempel his champion and his voice. The artworks therefore also present the adventures of a hero intertwined with the portrait of an "artiste maudit", very much along classical themes but presented in a very original framework.

Almost from publication, the 1964 book was considered by art critics and art historians as a masterpiece in Ernst's printed works, along with his two collage books *La femme 100 têtes* (1929) and *Une semaine de bonté* (1934). This interest has not declined: in December 2014, a copy of *Maximiliana* was sold at an auction in Artcurial for 129 900€, nearly four times its initial estimated value! However, only a few research articles in art history have investigated the project, and no trace of it can be found in astronomical circles. This article tries to fill this gap, showing that art can also shed an interesting light on the history of astronomy.

*Acknowledgments:* The author thanks S. Bianchi, E. Mossoux, G. Rauw, J. Schmitt and the Hamburg libraries, J. Cuypers and the ORB librarian, as well as A. Heward for their help and interesting discussions. She also wishes to thank the referees for comments that helped improving the paper.